%% file: main.tex
\documentclass{article}
\usepackage{spconf,amsmath,graphicx}
\usepackage{url}
\usepackage{siunitx}
\usepackage{amsfonts} 
\usepackage{amsmath}

\DeclareMathOperator*{\argmin}{arg\,min}

\usepackage{relsize}
\usepackage[printonlyused,withpage]{acronym} 

\newcommand{\expectOp}{\mathbb{E}}

\def\beamin{{\mathbf x}}
\def\beaminnum{{X}}
\def\weight{{\mathbf w}}
\def\beamout{{\mathbf y}}
\def\beamoutnum{{Y}}
\def\t{{\tau}}
\def\f{{f}}
\def\mic{{q}}
\def\TF{{(\t,\f)}}
\def\nMic{{Q}}
\def\nModel{{M}}
\def\nFreq{{F}}
\def\eigvec{{\mathbf U}}
\def\eigval{{\mathbf \Sigma}}
\def\NCM{{\mathbf R}}
\def\ATF{{\mathbf d}}
\def\steervec{{\mathbf a}}
\def\outarray{{\mathbf Z}}
\def\doa{{\Omega}}
\def\az{{\varphi}}
\def\inc{{\theta}}

\title{Subspace Hybrid Beamforming for head-worn microphone arrays}
\name{\parbox{\linewidth}{\centering%
Sina Hafezi\,$^{\text{\textdagger}}$, %
Alastair H. Moore\,$^{\text{\textdagger}}$, %
Pierre Guiraud\,$^{\text{\textdagger}}$, %
Patrick A. Naylor\,$^{\text{\textdagger}}$, \\%
Jacob Donley\,$^{\text{\textdaggerdbl}}$, %
Vladimir Tourbabin\,$^{\text{\textdaggerdbl}}$, %
Thomas Lunner\,$^{\text{\textdaggerdbl}}$} 
}

\address{$^{\text{\textdagger}}$ Electrical and Electronic Engineering, Imperial College London, London, UK\\
	 $^{\text{\textdaggerdbl}}$ Meta Reality Labs Research, Redmond, Washington,
	 USA}

\begin{document}

\newcommand{\acro}{\acrodef}\newcommand{\acroindefinite}{\acrodefindefinite}\input{sapacronyms.txt} 

%
\maketitle
\begin{abstract}
A two-stage multi-channel speech enhancement method is proposed which consists of a novel adaptive beamformer, Hybrid \ac{MVDR}, Isotropic-\ac{MVDR} (Iso), and a novel multi-channel spectral \ac{PCA} denoising. In the first stage, the Hybrid-\ac{MVDR} performs multiple \acp{MVDR} using a dictionary of pre-defined noise field models and picks the minimum-power outcome, which benefits from the robustness of signal-independent beamforming and the performance of adaptive beamforming. In the second stage, the outcomes of Hybrid and Iso are jointly used in a two-channel \ac{PCA}-based denoising to remove the `musical noise' produced by Hybrid beamformer. On a dataset of real `cocktail-party' recordings with head-worn array, the proposed method outperforms the baseline superdirective beamformer in noise suppression (fwSegSNR, SDR, SIR, SAR) and speech intelligibility (STOI) with similar speech quality (PESQ) improvement.
\end{abstract}

\input{body}


\vfill\pagebreak


\bibliographystyle{bib_all/IEEEbib_mod}
\bibliography{bib_all/sapstrings, bib_all/sapref, bib_all/local, bib_all/tmp}

\end{document}

%% file: sapacronyms.txt
\acro{3GPP}{3rd Generation Partnership Project}
\acro{a-SLAM}[aSLAM]{Acoustic \aclu{SLAM}}
\acro{aSLAM}{Acoustic \aclu{SLAM}}
\acro{A-SNR}{A-weighted \acs{SNR}}
\acro{AAC}{Advanced Audio Coding\acroextra{. A lossy codec used for digital audio.}}
\acro{AAD}{Auditory Attention Detection}
\acro{AAS}{American Auditory Soc.}
\acro{AASP}{Audio and Acoustic Signal Processing}
\acro{ABC}{Analytical with or without Bias Compensation}
\acro{ABR}{Auditory-Brainstem Response}
\acro{ACAWD}{Archivable Core Actual-Word Database}
\acro{ACB}{Adaptive Codebook}
\acro{ACC}{Accuracy}
\acro{ACE}{Acoustic Characterization of Environments\acroextra{. A noisy reverberant speech corpus and IEEE challenge run by the SAP group at Imperial College}}
\acro{ACELP}{Algebraic Code-Excited Linear Prediction}
\acro{ACF}{Autocorrelation Function}
\acro{ACK}{Acknowledgement}
\acro{ACL}{Access Control List}
\acro{ACR}{Absolute Category Rating}
\acro{AD}{Audio Diarization}
\acro{ADC}{Analogue-to-Digital Converter}
\acro{ADM}{Adaptive Differential Microphone}
\acro{ADPCM}{Adaptive Differential Pulse Code Modulation}
\acro{ADSL}{Asymmetric Digital Subscriber Line}
\acro{AE}{Almost Everywhere}
\acro{AES}{Audio Engineering Society}
\acro{AES2}[AES]{Advanced Encryption Standard}
\acro{AGC}{Automatic Gain Control}
\acro{AH}{Amplitude Histogram}
\acro{AI}{Articulation Index}
\acro{AI2}[AI]{Artificial Intelligence}
\acro{AI3}[AI]{Audio Inpainting}
\acro{AIC}{Akaike Information Criterion}
\acro{AIFF}{Audio Interchange File Format}
\acro{AIR}{Acoustic Impulse Response}
\acro{AIR2}[AIR]{Aachen Impulse Response}
\acro{AIRD}{Aachen Impulse Response Database}
\acro{ALC}{Automatic Level Control}
\acro{ALCons}{Articulation Loss of Consonants}
\acro{AM}{Amplitude Modulation}
\acro{AMDF}{Average Magnitude Difference Function\acroextra{. A function with similar properties to the cross- or autocorrelation but that requires no multiplication to evaluate.}}
\acro{AMR}{Adaptive Multi-Rate}
\acro{AMR-NB}{Adaptive Multi-Rate Narrow Band}
\acro{AMR-WB}{Adaptive Multi-Rate Wide Band}
\acro{AMS}{Amplitude Modulation Spectrogram}
\acro{ANC}{Adaptive Noise Canceller}
\acro{ANS}{Autocorrelation-Based Noise Subtraction}
\acro{ANSI}{American National Standards Institute}
\acro{ANU}{Australian National University}
\acro{APLAWD}{Archivable Priority List Actual-Word Database}
\acro{AR}{Autoregressive}
\acro{ARD}{Arbeitsgemeinschaft der \"{o}ffentlich-rechtlichen Rundfunkanstalten der Bundesrepublik Deutschland\acroextra{. `Consortium (``Working group'') of the public-law broadcasting institutions of the Federal Republic of Germany'}}
\acro{ARMA}{Autoregressive Moving Average}
\acro{AS}{Audio Segmentation}
\acro{AS2}[AS]{Almost Surely}
\acro{ASA}{Acoustic Scene Analysis}
\acro{ASIO}{Audio Stream Input/Output\acroextra{. A computer soundcard protocol with low latency developed by Steinberg.}}
\acro{ASK}{Amplitude Shift Keying}
\acro{ASLP}{Audio, Speech, and Language Processing}
\acro{ASM}{Acoustic Scene Mapping}
\acro{ASR}{Automatic Speech Recognition}
\acro{ASS}{Approximate Spectrum Substitution}
\acro{AST}{Acoustic Source Tracking}
\acro{AST2}[AST]{Asymmetric Sampling in Time}
\acro{ATF}{Acoustic Transfer Function\acroextra{. The Fourier Transform of the \acs{RIR}.}}
\acro{ATLM}{Acoustic Tokenization and Language Modelling}
\acro{ATR}{Advanced Telecommunications Research Institute International\acroextra{, Kyoto, Japan}}
\acro{AUC}{Area under the curve}
\acro{AURORA}{Aurora Experimental Framework for the Evaluation of the Performance of Speech Recognition Systems under Noisy Conditions}
\acro{AUV}{Autonomous Underwater Vehicle}
\acro{AV}{Audio-Visual}
\acro{AWGN}{Additive White Gaussian Noise}
\acro{AWS}{Approximate Waveform Substitution}
\acro{BASIE}{Bayesian Adaptive Speech Intelligibility Estimation}
\acro{BCE}{Blind Channel Estimation}
\acro{BCL}{Bekesy Comfortable Loudness}
\acro{BCR}{Block-Coordinate Relaxation}
\acro{BEM}{Boundary Element Method}
\acro{BER}{Bit Error Rate}
\acro{BIBO}{Bounded-Input Bounded-Output}
\acro{BIC}{Bayesian Information Criterion}
\acro{Bk}{Berksons\acroextra{. A unit for measuring intelligibility.}}
\acro{BK}[B\&K]{Br{\"u}el and Kj{\ae}r}
\acro{BLSTM}{Bidirectional \acs{LSTM}}
\acro{BM}{Blocking Matrix}
\acro{BO-SCPHD}{Bearing-only \acs{SC-PHD}}
\acro{BO-SLAM}{Bearing-only \acs{SLAM}}
\acro{BOT}{Bearing-only tracking}
\acro{BP}{Basis Pursuit}
\acro{BPCC}{Basis Pursuit with Clipping Constraints}
\acro{BPDN}{Basis Pursuit Denoising}
\acro{BPM}{Beats Per Minute}
\acro{BPSK}{Binary Phase Shift Keying}
\acro{BR}{Barrodale and Roberts' (algorithm)}
\acro{BRI}{Basic Rate Index}
\acro{BSD}{Bark Spectral Distortion}
\acro{BSI}{Blind System Identification}
\acro{BSIM}{Binaural Speech Intelligbility Model}
\acro{BSS}{Blind Source Separation}
\acro{BSTOI}{Binaural \acs{STOI}}
\acro{BW}{Bandwidth}
\acro{BZ}{Back-to-Zero}
\acro{C4DM}{Centre for Digital Music}
\acro{C50}[$C_\textrm{50}$]{Clarity Index}
\acro{C-GFB}{Combination Gas-fired Boiler}
\acro{CART}{Classification and Regression Tree}
\acro{CASA}{Computational Auditory Scene Analysis}
\acro{CBR}{Constant Bit Rate}
\acro{CCC}{Cross-Correlation Coefficient}
\acro{CCCC}{DARPA CSR Corpus Coordinating Committee}
\acro{CCD}{Charge-Coupled Device}
\acro{CCI}{Call Clarity Index}
\acro{CCITT}{Consultative Committee for International Telephony and Telegraphy}
\acro{CCM}{Contralateral Competing Message}
\acro{CCR}{Comparison Category Rating}
\acro{CDB}{Constant Directivity Beamformer}
\acro{CDMA}{Code Division Multiple Access}
\acro{CELP}{Code-excited Linear Prediction}
\acro{CHIEF}{Combined Helmholtz Integral Equation Formulation}
\acro{CIT}{Constrained Initial Taps}
\acro{CL}{Clipping Level}
\acro{CLEAR}{Centre for Law Enforcement Audio Research}
\acro{CLID}{Cluster Identification Test}
\acro{CLT}{Central Limit Theorem}
\acro{CMA}{Constant Modulus Algorithm}
\acro{CMASI}{Coherence Modulated Acoustic Speckle Interferometry}
\acro{CMB}{Cosmic Microwave Background}
\acro{CNC}{Consonant-Nucleus-Consonant}
\acro{CNG}{Comfort Noise Generation}
\acro{CODEC}{Coder-Decoder}
\acro{CPE}{Customer Premises Equipment}
\acro{CPHD}{Cardinalized \acs{PHD}}
\acro{CRACD}{Codec-robust Automatic Clipping Detector}
\acro{CRC}{Cyclic Redundancy Check}
\acro{CS}{Channel Shortening}
\acro{CS2}[CS]{Compressive Sensing}
\acro{CSP}{Communications and Signal Processing}
\acro{CSR-WSJ}{Continuous Speech Recognition Wall Street Journal Phase 1\acroextra{ database}}
\acro{CST}{Connected Speech Test\acroextra{ speech corpus}}
\acro{CT}{Conversation Test}
\acro{CTTN}{Comparative Tolerance to Noise}
\acro{CV}{Coefficient of Variation}
\acrodefplural{CV}{Coefficients of Variation}
\acro{CV2}[CV]{Constant Velocity}
\acro{CVC}{Consonant-Vowel-Consonant}
\acro{CW}{Continuous Wave}
\acro{CWM}{Centre-Weighted Median}
\acro{CWMY}{Centre-Weighted Myriad}
\acro{CWT}{Continuous Wavelet Transform}
\acro{DAC}{Digital-to-Analogue Converter}
\acro{DAM}{Diagnostic Acceptability Measure}
\acro{DAQ}{Data Acquisition}
\acro{DARPA}{Defense Advanced Research Projects Agency\acroextra{ of the United States Dept. of Defense}}
\acro{DARPA-RMD}{\acs{DARPA} 1000-Word Resource Management Database\acroextra{ for Continuous Speech Recognition}}
\acro{DAW}{Digital Audio Workstation}
\acro{dB}{Decibel}
\acro{dBFS}{\acs{dB} Full Scale}
\acro{DBN}{Deep Belief Network}
\acro{DBSTOI}{Deterministic \acs{BSTOI}}
\acro{DC}{Direct Current}
\acro{DCME}{Digital Circuit Multiplexing Equipment}
\acro{DCR}{Degradation Category Rating}
\acro{DCT}{Discrete Cosine Transform}
\acro{DDR}{Direct-to-diffuse ratio}
\acro{DDR3}{Double Data Rate Type Three}
\acro{DECT}{Digital European Cordless Telecommunication}
\acro{DeLILAH}{Detection of Clipping using Least Squares Residuals and Iterated Logarithm Amplitude Histogram}
\acro{DENBE}{\acs{DRR} Estimation using a Null-Steered Beamformer}
\acro{DET}{Detection Error Trade-off}
\acro{Dev}{Development\acroextra{ dataset of the \acs{ACE} Challenge}}
\acro{DFT}{Discrete Fourier Transform}
\acro{DI}{Directivity Index}
\acro{DirectX}{\acroextra{A programming interface developed by Microsoft for handling tasks related to multimedia.}}
\acro{DIRHA}{Distant-speech Interaction for Robust Home Applications\acroextra{ multi-microphone multi-language acoustic speech corpus}}
\acro{DMA}{Differential Microphone Array}
\acro{DMT}{Discrete Multi-Tone}
\acro{DMV}{Dynamically Managed Voice\acroextra{ system}}
\acro{DNN}{Deep Neural Network}
\acro{DNR}{Dynamic Noise Reduction}
\acro{DoA}{Direction-of-Arrival}
\acrodefplural{DoA}[DoAs]{Directions-of-Arrival}
\acro{DOA}{Direction-of-Arrival}
\acrodefplural{DOA}[DOAs]{Directions-of-Arrival}
\acro{DP}{Dynamic Programming}
\acro{DPCM}{Differential Pulse Code Modulation}
\acro{DPD}{Direct-Path Dominance}
\acro{DPD-MUSIC}{Direct-Path Dominance Multiple Signal Classification}
\acro{DR}{Douglas-Rachford}
\acro{DR2}{Dynamic Range}
\acro{DRM}{Diagnostic Rhyme Test}
\acro{DRR}{Direct-to-Reverberant Ratio}
\acro{DRNN}{Deep Recurrent Neural Net}
\acro{DRT}{Diagnostic Rhyme Test}
\acro{DSB}{Delay-and-Sum Beamformer}
\acro{DSOBM}{Deterministic \acs{SOBM}}
\acro{DSP}{Digital Signal Processing}
\acro{DSPS}{Double Sides Periodic Substitution}
\acro{DSR}{Distributed Speech Recognition}
\acro{DSWS}{Double Sides Waveform Substitution}
\acro{DTMF}{Dual Tone Multi-Frequency}
\acro{DTX}{Discontinued Transmission}
\acro{DWT}{Discrete Wavelet Transform}
\acro{EARS}{Embodied Audition for RobotS}
\acro{EBF}{Eigen-beamformer}
\acro{EBU}{European Broadcasting Union}
\acro{EC}{Echo Canceller}
\acro{EDC}{Energy Decay Curve}
\acro{EDR}{Energy Decay Relief}
\acro{EDF}{Energy Decay Function}
\acro{EEG}{Electroencephalography}
\acro{EER}{Equal Error Rate}
\acro{EFICA}{Efficient Fast Independent Component Analysis}
\acro{EIR}{Equalized Impulse Response}
\acro{EKF}{Extended Kalman Filter}
\acro{EKF-SLAM}[EKF-SLAM]{\acs{EKF} \acs{SLAM}}
\acro{EKF-SLAM2}[EKF-SLAM]{Extended Kalman Filter \acs{SLAM}}
\acro{EL}{Echo Loss}
\acro{ELF}{Extremely Low Frequency}
\acro{ELRA}{European Languages Research Association}
\acro{EM}{Estimation-Maximization\acroextra{. An iterative technique to solve certain optimization problems.}}
\acro{EMA}{Exponential Moving Average}
\acro{EMIB}{Eigenmike Microphone Interface Box}
\acro{ENF}{Electrical Network Frequency}
\acro{EPSRC}{Engineering and Physical Sciences Research Council}
\acro{EQ}{Equalisation}
\acro{ERB}{Equivalent Rectangular Bandwidth}
\acro{ERP}{Ear Reference Point (cf. ITU-T Rec. P.64 1999)}
\acro{ESA}{Early Stage Assessment}
\acro{ESM}{Equivalent Source Method}
\acro{ESPRIT}{Estimation of Signal Parameters via Rotational Invariance Techniques}
\acro{ETAN}{Equivalent Tolerance to Additional Noise} 
\acro{ETSI}{European Telecommunications Standards Institute}
\acro{EURASIP}{European Association for Signal Processing}
\acro{EUSIPCO}{European Signal Processing Conference}
\acro{Eval}{Evaluation\acroextra{ dataset of the \acs{ACE} Challenge}}
\acro{EVD}{EigenValue Decomposition}
\acro{F1}{F1 Score}
\acro{FAR}{False Alarm Rate}
\acro{FastSLAM}[FastSLAM]{FActored Solution To Simultaneous Localization and Mapping}
\acro{FastSLAM2}[FastSLAM]{FActored Solution To \acs{SLAM}} 
\acro{FAU}{Friedrich-Alexander-Universit{\"a}t}
\acro{FB}{Forward-Backward}
\acro{FB2}[FB]{Fullband}
\acro{FBF}{Fixed Beamformer}
\acro{FBSS}{Forward-Backward Spatial Smoothing}
\acro{FCC}{Federal Communications Commission}
\acro{FDM}{Frequency Division Multiplexing}
\acro{FDR}{False Discovery Rate}
\acro{FDR2}[FDR]{Free-Decay Region}
\acro{FEC}{Forward Error Correction}
\acro{FEM}{Finite Element Method}
\acro{FFI}{Norwegian Defence Research Establishment}
\acro{FFT}{Fast Fourier Transform}
\acroindefinite{FFT}{an}{a}
\acro{FIFO}{First-In First-Out}
\acro{FIR}{Finite Impulse Response\acroextra{. A filter whose output is a weighted sum of past input values and whose system function contains only zeros and no poles.}}
\acro{FISM}{Fast Image Source Method}
\acro{FISST}{Finite Set STatistics}
\acro{FLOM}{Fractional Lower-Order Moments}
\acro{FLOS}{Fractional Lower-Order Statistics}
\acro{FM}{Frequency Modulation}
\acro{FMM}{Fast Multipole Method}
\acro{FN}{False Negative}
\acro{FN2}{Nth Formant}
\acro{FNR}{False Negative Rate}
\acro{FORTRAN}{The IBM Mathematical Formula Translating System}
\acro{FoV}{Field of View}
\acro{FP}{False Positive}
\acro{FPR}{False Positive Rate}
\acro{FPS}{Frames Per Second}
\acro{FRI}{Finite Rate of Innovation}
\acro{FSB}{Filter-and-Sum Beamformer}
\acro{FSK}{Frequency Shift Keying}
\acro{FT}{Flat-Top}
\acro{FWER}{Familywise Error Rate}
\acro{FWSSNR}{Frequency-Weighted Segmental \acs{SNR}}
\acro{FWSSRR}{Frequency-Weighted \acs{SSRR}}
\acro{G.711}{\acs{PCM} of Voice Frequencies}
\acro{GARCH}{Generalized Auto-regressive Conditional Heteroscedasticity}
\acro{GBW}{Gain Bandwidth Product}
\acro{GCC}{Generalized Cross-Correlation}
\acro{GCC-PHAT}{Generalized Cross-Correlation with Phase Transform\acroextra{ method of estimating \acs{TDoA}}}
\acro{GCF}{Global Coherence Field}
\acro{GGD}{Generalized Gaussian Distribution}
\acro{GGD2}[G$\Gamma$D]{Generalised Gamma Distribution}
\acro{GM}{Gaussian Mixture}
\acro{GM-PHD}{Gaussian Mixture \acs{PHD}}
\acro{GMCA}{Generalized Morphological Component Analysis}
\acro{GMM}{Gaussian Mixture Model\acroextra{. An approximation to an arbitrary probability density function that consists of a weighted sum of Gaussian distributions}}
\acro{GNSS}{Global Navigation Satellite System}
\acro{GPRS}{General Packet Radio Services}
\acro{GPS}{Global Positioning System}
\acro{GSC}{Generalized Sidelobe Canceller}
\acro{GSM}{Global System for Mobile Communications}
\acro{GSM-EFR}{\acs{GSM} Enhanced Full Rate Codec}
\acro{GSM-FR}{\acs{GSM} Full Rate Codec}
\acro{GSM-HR}{\acs{GSM} Half Rate Codec}
\acro{GUI}{Graphical User Interface}
\acro{HAAC}{High Amplitude Audio Capture}
\acro{HATS}{Head and Torso Simulator}
\acro{HERB}{Harmonicity-based dEReverBeration}
\acro{HFT}{Hands-Free Terminal}
\acro{HI}{Hearing-Impaired}
\acro{HIE}{Helmholtz Integral Equation}
\acro{HISAS}{High resolution Interferometric \ac{SAS}}
\acro{HINT}{Hearing-in-Noise Test}
\acro{HLT}{Human Language Technology}
\acro{HMM}{Hidden Markov Model}
\acro{HOS}{Higher-Order Statistics}
\acro{HPF}{High-Pass Filter}
\acro{HR}{Half Rate (\acs{GSM} Codec)}
\acro{HRI}{Human-Robot Interaction}
\acro{HRTF}{Head-related Transfer Function}
\acro{HSA}{Hearing, Speech, Audio\acroextra{ technology group at Fraunhofer IDMT}}
\acro{HSD}{Hybrid Steepest Descent}
\acro{HT}{Hannan-Thomson}
\acro{HTK}{Hidden Markov Model Tool Kit}
\acro{IBM}{Ideal Binary Mask}
\acro{IC}{Interference Canceller}
\acro{ICA}{Independent Component Analysis}
\acro{ICASSP}{Intl. Conf. on Acoustics, Speech and Signal Processing}
\acro{ID}{Identifier}
\acro{IDMT}{Institute for Digital Media Technology}
\acro{iDEN}{Integrated Digital Enhanced Network}
\acro{IEC}{International Electrotechnical Commission}
\acro{IEEE}{Institute of Electrical and Electronics Engineers}
\acro{IET}{Institute of Engineering and Technology}
\acro{IETF}{Internet Engineering Task Force}
\acro{IFFT}{Inverse Fast Fourier Transform}
\acro{IHC}{Inner Hair Cell}
\acro{IHT}{Iterative Hard Thresholding}
\acro{IID}{Independent and Identically Distributed}
\acro{IIR}{Infinite Impulse Response\acroextra{. A filter whose output is a weighted sum of both past input and past output values and whose system function contains both poles and zeros.}}
\acro{IL}{Iterated Logarithm\acroextra{, the logarithm of the logarithm}}
\acro{ILAH}{Iterated Logarithm Amplitude Histogram\acroextra{ clipping detection method}}
\acro{ILD}{Interaural Level Difference}
\acro{IMCRA}{Improved Minima Controlled Recursive Averaging\acroextra{. A technique for blindly estimating the spectrum of additive noise in a signal.}}
\acro{IMD}{Inter-Modulation Distortion}
\acro{IMSI}{International Mobile Subscriber Identity}
\acro{IMU}{Inertial Measurement Unit}
\acro{INMD}{In-service Non-intrusive Measurement Device}
\acro{INTERSPEECH}{Annual Conference of the \acs{ISCA}}
\acro{IO}{Infinitely Often}
\acro{IP}{Internet Protocol}
\acro{IPA}{International Phonetic Association}
\acro{IPD}{Interaural Phase Difference}
\acro{IRM}{Ideal Ratio Mask}
\acro{IRS}{Inverse repeated Sequence\acroextra{. A pseudo random sequence used for impulse response measurement.}}
\acro{IRS2}[IRS]{Intermediate Reference System}
\acro{IS}{Importance Sampling}
\acro{ISCA}{International Speech Communication Association}
\acro{ISDN}{Integrated Services Digital Network}
\acro{ISFT}{Inverse \acl{SFT}}
\acro{ISO}{Intl. Organization for Standardization}
\acro{ISP}{Intensity Spectral Profile}
\acro{IST}{Iterative Soft Thresholding}
\acro{ISTFT}{Inverse Short Time Fourier Transform}
\acro{ISVR}{Institute of Sound and Vibration Research\acroextra{, Southampton University, UK}}
\acro{ITD}{Interaural Time Difference}
\acro{ITF}{Interaural Transfer Function}
\acro{ITU}{International Telecommunication Union}
\acro{ITUR}[ITU-R]{International Telecommunication Union\acroextra{ Radiocommunication Sector}}
\acro{ITUT}[ITU-T]{International Telecommunication Union\acroextra{ Telecommunication Standardisation Sector}}
\acro{IUWT}{Isotropic Undecimated Wavelet (Starlet) Transform}
\acro{IWAENC}{Intl. Workshop on Acoustic Signal Enhancement}
\acro{IWAENC_PRE_2012}[IWAENC]{Intl. Workshop Acoustic Echo and Noise Control}
\acro{IWASE}{Intl. Workshop on Acoustic Signal Enhancement}
\acro{JADE}{Joint Approximate Diagonalization of Eigen-Matrices}
\acro{JPDA}{Joint Probabilistic Data Association}
\acro{JPEG}{Joint Photographic Experts Group}
\acro{KDE}{Kernel Density Estimate}
\acro{KF}{Kalman Filter}
\acro{KFD}{Kernel Fisher Discriminant\acroextra{ Analysis}}
\acro{KL}{Karhunen-Lo{\'{e}}ve}
\acro{KLT}{Karhunen-Lo{\'{e}}ve Transform}
\acro{KST}{Kolmogorov-Smirnov Test}
\acro{LAD}{Least Absolute Deviation}
\acro{LAN}{Local Area Network}
\acro{LARS}{Least Angle Regression}
\acro{LAT}[L$_{\textrm{AT}}$]{Equivalent Continuous Sound Level\acroextra{. Also called Leq}}
\acro{LBR}{Low Bitrate Redundancy}
\acro{LC}{Local Criterion}
\acro{LCMP}{Linearly Constrained Minimum Power}
\acro{LCMV}{Linearly Constrained Minimum Variance}
\acro{LCWM}{Linear Combination of Weighted Medians}
\acro{LDC}{Linguistic Data Consortium}
\acro{LEM}{Loudspeaker-Enclosure-Microphone System}
\acro{Leq}[L$_{\textrm{eq}}$]{Equivalent Continuous Sound Level\acroextra{. Also called LAT}}
\acro{LF}{Liljencrants-Fant\acroextra{. The developers of a glottal waveform model}}
\acro{LHS}{Left-Hand Side}
\acro{LID}{Language Identification}
\acro{LILAH}{\acs{LSR2}-\acs{ILAH}\acroextra{ clipping detection method}}
\acro{LIME}{LInear Predictive Multi-input Equalization\acroextra{ algorithm}}
\acro{LiNoPS}{Lightweight Noise Protection System}
\acro{LLN}{Law of Large Numbers}
\acro{LLR}{Log-Likelihood Ratio}
\acro{LLS}{Logarithmic Least Squares}
\acro{LMA}{Least Mean Absolute}
\acro{LMS}{Least Mean Squares\acroextra{ adaptive filter}}
\acro{LNA}{Low Noise Amplifier}
\acro{LoS}{Line-of-Sight}
\acro{LOT}{Listening-Only Test}
\acro{LP}{Linear Parameter}
\acro{LP2}[LP]{Linear Predictive}
\acro{LP3}[LP]{Linear Prediction}
\acro{LPC}{Linear Predictive Coding\acroextra{. An autoregressive model of speech production.}}
\acro{LQO}{Listening Quality Objective}
\acro{LREC}{Conf. on Language Resources and Evaluation}
\acro{LS}{Least Squares}
\acro{LSA}{Log Spectral Amplitude}
\acro{LSB}{Lower Side-Band}
\acro{LSB2}[LSB]{Least Significant Bit}
\acro{LSD}{Log Spectral Distortion}
\acro{LSP}{Line Spectrum Pairs}
\acro{LSR}{Late Stage Review}
\acro{LSR2}[LSR]{Least Squares Residuals\acroextra{ clipping detection method}}
\acro{LSRT}{Least Squares Residuals with Thresholding\acroextra{ clipping detection method}}
\acro{LSTM}{Long Short-Term Memory}
\acro{LTASS}{Long Term Average Speech Spectrum}
\acro{LTI}{Linear Time Invariant}
\acro{LTP}{Long Term Prediction}
\acro{LU}{Loudness Unit}
\acro{LUFS}{Loudness Units Full-Scale}
\acro{MA}{Moving Average}
\acro{MAC}{Multiply Accumulate Operation}
\acro{MAD}{Median Absolute Deviation}
\acro{MAE}{Mean Absolute Error}
\acro{MAP}{Maximum \emph{a posteriori}}
\acro{MARDY}{Multichannel Acoustic Reverberation Database at York}
\acro{MARS}{Multivariate Adaptive Regression Splines}
\acro{MBF}{Matched Filter Beamformer}
\acro{MC}{Monte Carlo}
\acro{MCA}{Morphological Component Analysis}
\acro{MCC}{Matthew's Correlation Coefficient}
\acro{MCEQ}{MultiChannel EQualisation}
\acro{MCS}{Multidimensional Colouration Space}
\acro{MDCT}{Modified Discrete Cosine Transform}
\acro{MDL}{Minimum Description Length}
\acro{MDS}{Multidimensional Scaling}
\acro{Mel}{\acroextra{A non-uniform frequency scale corresponding to perceived frequency. It is approximately linear at low frequencies and logarithmic at high frequencies.}}
\acro{MELP}{Mixed Excitation Linear Prediction}
\acro{MFCC}{Mel-frequency Cepstral Coefficients}
\acro{MFS}{Method of Fundamental Solutions}
\acro{MFSK}{Multi-Frequency Shift Keying}
\acro{MHT}{Multi-Hypotheses Tracking}
\acro{MI}{Mutual Information}
\acro{MIMO}{Multiple-Input-Multiple-Output}
\acro{MINT}{Multiple-input/output INverse Theorem}
\acro{MIRS}{Motorola Integrated Radio System}
\acro{MISO}{Multiple-Input-Single-Output}
\acro{MIT}{Massachusetts Institute of Technology}
\acro{MIT-LCS}{Massacchusetts Institute of Technology Laboratory for Computer Science}
\acro{ML}{Machine Learning}
\acro{MLE}{Maximum Likelihood Estimation}
\acro{ML-TDoA}{Maximum Likelihood Time Difference of Arrival}
\acro{MLD}{Masking Level Difference}
\acro{MLMF}{Machine Learning with Multiple Features}
\acro{MLS}{Maximum Length Sequence\acroextra{ of pseudo random bits.}}
\acro{MMSE}{Minimum Mean Squared Error}
\acro{MMT}{Multiscale Median Transform}
\acro{MNRU}{Modulated Noise Reference Unit}
\acro{MOM}{Mean of Maximum}
\acro{MOS}{Mean Opinion Score}
\acro{MOS-LQO}{Mean Opinion Score - Listening Quality Objective}
\acro{MP}{Matching Pursuit}
\acro{MP3}{\acs{MPEG}-2 Audio Layer III}
\acro{MPDR}{Minimum Power Distortionless Response\acroextra{ beamformer}}
\acro{MPEG}{Moving Picture Experts Group}
\acro{MRF}{Markov Random Field}
\acro{MRP}{Mouth Reference Point (cf. ITU-T Rec. P.64 1999)}
\acro{MRT}{Modified Rhyme Test}
\acro{MS}{Minimum Statistics}
\acro{MSB}{Most Significant Bit}
\acro{MSC}{Mean Square Coherence}
\acro{MSE}{Mean Square Error}
\acro{MSN}{Multiple Subscriber Number}
\acro{MSNR}{Maximum \acs{SNR}}
\acro{MTF}{Modulation Transfer Function}
\acro{MTM}{Modified Trimmed Mean}
\acro{MUSCLE}{MeasUred Single-CLustEr}
\acro{MUSHRA}{Multi-stimuli Test with Hidden Reference and Anchor}
\acro{MUSHRAR}{Multi-stimuli Test with Hidden Reference and Anchor for Reverberation}
\acro{MUSIC}{Multiple Signal Classification}
\acro{MVDR}{Minimum Variance Distortionless Response\acroextra{ beamformer}}
\acro{MWF}{Multi-channel Wiener Filter}
\acro{NAH}{Nearfield Acoustic Holography}
\acro{NB}{Narrowband}
\acro{NCM}{Noise Covariance Matrix}
\acro{NH}{Normal-Hearing}
\acro{NIRA}{Non-Intrusive Room Acoustics}
\acro{NISE}{Non-Intrusive \acs{SNR} estimation}
\acro{NISI}{Non-Intrusive Speech Intelligibility Estimation}
\acro{NISQ}{Non-Intrusive Speech Quality Estimation}
\acro{NIST}{National Institute of Standards and Technology}
\acro{NL}{Noise Level}
\acro{NLA}{Non-Linear Approximation}
\acro{NLMS}{Normalized Least Mean Squares\acroextra{ adaptive filter}}
\acro{NMCFLMS}{Normalized Multichannel Frequency Domain Least Mean Square}
\acro{NMF}{Non-negative Matrix Factorization}
\acro{NOISEX-92}{Database to Study the Effect of Additive Noise on Speech Recognition Systems}
\acro{NOIZEUS}{Noisy Speech Corpus for Evaluation of Speech Enhancement Algorithms}
\acro{NOSRMR}{Normalized Overall \acs{SRMR}}
\acro{NOS}{Number of Sources}
\acro{NOSRMR}{Normalized Overall \acs{SRMR}}
\acro{NPM}{Normalized Projection Misalignment}
\acro{NPV}{Negative Predictive Value}
\acro{NR}{Noise Reduction}
\acro{NS}{Noise Suppression}
\acro{NSRMR}{Normalised \acs{SRMR}}
\acro{NSRR}{Normalized Signal-to-Reverberation Ratio}
\acro{NSRMR}{Normalised \acs{SRMR}}
\acro{NSV}{Negative-Side Variance}
\acro{NTP}{Network Time Protocol}
\acro{NV}{Noise-Vocoding}
\acro{OBL}{Octave Band Level}
\acro{ODF}{Overdrive Factor}
\acro{Ofcom}{Office of Communications\acroextra{, the independent regulator and competition authority for the UK communications industries}}
\acro{OFDM}{Orthogonal Frequency Division Multiplexing}
\acro{OHC}{Outer Hair Cell}
\acro{OIM}{Objective Intelligibility Measure}
\acro{OLA}{Overlap-add}
\acro{OM-LSA}{Optimally Modified Log-Spectral Amplitude{ Estimator}}
\acro{OMP}{Orthogonal Matching Pursuit}
\acro{OSI}{Open Systems Interconnection}
\acro{OSPA}{Optimal Subpattern Assignment}
\acro{OSRMR}{Overall \acs{SRMR}}
\acro{PAB-SRMR}{Per acoustic band \acs{SRMR}}
\acro{PALM}{Passive Acoustic Localization and Mapping}
\acro{PAMS}{Perceptual Analysis Measurement System}
\acro{PARCOR}{Partial Correlation Coefficients}
\acro{PB}{Phonetically Balanced}
\acro{PBF}{Positive Boolean Function}
\acro{PCA}{Principal Components Analysis}
\acro{PCM}{Pulse-Code Modulation}
\acro{PDA}{Personal Digital Assistant}
\acro{PDA2}[PDA]{Probabilistic Data Association}
\acro{PDE}{Partial Differential Equation}
\acro{pdf}{Probability Density Function}
\acro{PDF}{Probability Density Function}
\acro{PE}{Parameter Estimation}
\acro{PEASS}{Perceptual Evaluation for Audio Source Separation}
\acro{PEFAC}{Pitch Estimation Filter with Amplitude Compression}
\acro{PESQ}{Perceptual Evaluation of Speech Quality}
\acro{PF}{Psychometric Function}
\acro{pgfl}[p.g.fl.]{Probability Generating Functional}
\acro{PHAT}{Phase Transform}
\acro{PHD}{Probability Hypothesis Density}
\acro{PIP}{Peak-Image Pairing}
\acro{PIV}{Pseudo-Intensity Vector}
\acro{PL}{Pseudo-likelihood}
\acro{PLC}{Packet Loss Concealment}
\acro{PLL}{Phase Locked Loop}
\acro{PLP}{Perceptual Linear Prediction}
\acro{PLR}{Perceived Level of Reverberation}
\acro{PM}{Phase Modulation}
\acro{PMF}{Probability Mass Function}
\acro{PMOS}{Predicted Mean Opinion Score}
\acro{POLQA}{Perceptual Objective Listening Quality Analysis}
\acro{POTS}{Plain Old Telephone Service}
\acro{PPP}{Poisson Point Process}
\acro{PPS}{Pulse-Per-Second}
\acro{PPV}{Positive Predictive Value}
\acro{PRLM}{Phoneme Recognition and Language Modelling}
\acro{PRP}{Pair-wise Relative Phase-ratio}
\acro{PSD}{Power Spectral Density}
\acro{PSK}{Phase Shift Keying}
\acro{PSNR}{Peak Signal-to-Noise Ratio}
\acro{PSOLA}{Pitch Synchronous Overlap Add\acroextra{. A method of scaling a signal in time and pitch independently.}}
\acro{PSQM}{Perceptual Speech Quality Measurement}
\acro{PSTN}{Public Switched Telephone Network}
\acro{PW}{Plane-Wave}
\acro{PWD}{Plane-Wave Decomposition}
\acro{PTA}{Pure-Tone Audiology}
\acro{QAM}{Quadrature Amplitude Modulation}
\acro{QILAH}{Quadrisected Iterated Logarithm Amplitude Histogram\acroextra{ clipping detection method}}
\acro{QMF}{Quadrature Mirror Filter}
\acro{QoE}{Quality-of-Experience}
\acro{QoS}{Quality-of-Service}
\acro{QPSK}{Quadrature Phase Shift Keying}
\acro{RADAR}{RAdio Detection And Ranging}
\acro{RASTA}{Relative Spectral}
\acro{RASTA-PLP}{Relative Spectral Perceptual Linear Prediction}
\acro{RASTI}{Room Acoustics Speech Transmission Index\acroextra{ (superseded by STIPA)}}
\acro{RBM}{Restricted Boltzmann Machine}
\acro{RB-PHD}{Rao-Blackwellised \acs{PHD}}
\acro{RBPF}{Rao-Blackwellised Particle Filter}
\acro{RC}{Relative Criterion}
\acro{RDT}[$R_\textrm{DT}$]{Reverberation Decay Tail}
\acro{RDTF}{Relative Direct Transfer Function}
\acro{RF}{Radio Frequency}
\acro{RFI}{Radio Frequency Interference}
\acro{RFS}{Random Finite Set}
\acro{RHS}{Right-Hand Side}
\acro{RIP}{Restricted Isometry Property}
\acro{RIR}{Room Impulse Response}
\acro{RLS}{Recursive Least Squares\acroextra{ adaptive filter}}
\acro{RLSD}{Relative Log Spectral Distortion}
\acro{RMCLS}{Relaxed Multichannel Least Squares}
\acro{RMCLS-CIT}{Relaxed MultiChannel Least-Squares with Constrained Initial Taps}
\acro{RMS}{Root Mean Square}
\acro{RMSE}{Root Mean Square Error}
\acro{RNN}{Recurrent Neural Net}
\acro{ROC}{Receiver Operating Characteristic}
\acro{ROHC}{Robust Header Compression}
\acro{RP}{Received Pronunciation}
\acro{RPE}{Regular Pulse Excitation}
\acro{RRTF}{Relative Real-Time Factor}
\acro{RS}{Reverberation Suppression}
\acro{RSM}{Reflector Source Method}
\acro{RSV}{Room Spectral Variance}
\acro{RT}{Reverberation Time}
\acro{RTAN}{Robustness to Additional Noise}
\acro{RTF}{Real-Time Factor}
\acro{RTF2}[RTF]{Room Transfer Function}
\acro{RTF3}[RTF]{Relative Transfer Function}
\acro{RV}{Random Variable}
\acro{RVP}{Recursive Vector Projection}
\acro{RWTH}{Rheinisch-Westf\"{a}lische Technische Hochschule}
\acro{S50}[$S_{50}$]{Intelligibility Function Gradient at the \ac{SRT}}
\acro{SA}{Spectral Amplitude}
\acro{SAA}{Synthetic Aperture Audio}
\acro{SAP}{Speech And Audio Processing}
\acro{SAR}{Speech-to-Artifact Ratio}
\acro{SAR2}[SAR]{Speaker Alternation Rate}
\acro{SAR3}[SAR]{Synthetic Aperture \ac{RADAR}}
\acro{SAS}{Synthetic Aperture \ac{SONAR}}
\acro{SB}{Subband}
\acro{SC-PHD}{Single Cluster \acs{PHD}}
\acro{SCAF}{Single Channel Adaptive Filter}
\acro{SCB}{Stochastic Codebook}
\acro{SCNR}{Single-Channel Noise Reduction}
\acro{SCOT}{Smoothed Coherence Transform}
\acro{SCM}{Sample Covariance Matrix}
\acro{SCNR}{Single-Channel Noise Reduction}
\acro{SC-PHD}{Single Cluster \acs{PHD}}
\acro{SCR}{Signal-to-Competition Ratio}
\acro{SCRIBE}{Spoken Corpus of British English}
\acro{SCT}{Speech Corruption Toolkit}
\acro{SCT2}{Short Conversation Test}
\acro{SD}{Semantic Differential}
\acro{SDB}{Superdirective Beamformer}
\acro{SDD}{Spectral Decay Distributions}
\acro{SDDMSB}{\acs{SDD} with Mel-spaced frequency bands}
\acro{SDDSA}{\acs{SDD} with Mel-spaced frequency bands and selective averaging}
\acro{SDDSA-G}{\acs{SDDSA} with Gerkmann noise estimator}
\acro{SDDSA-H}{\acs{SDDSA} with Hendriks noise estimator}
\acro{SDR}{Software Defined Radio}
\acro{SDR2}{Speech}
\acro{SDRAM}{Synchronous Dynamic Random Access Memory}
\acro{SDT}{Speech Description Taxonomy}
\acro{SEDF}{Subband \ac{EDF}}
\acro{SEMG}{Surface Electromyography}
\acro{SFDR}{Spurious Free Dynamic Range}
\acro{SFM}{Single Feature with Mapping}
\acro{SFT}{Spherical Fourier Transform}
\acro{SH}{Spherical Harmonic}
\acro{SHD}{Spectral Harmonic Decomposition}
\acro{SHD2}[SHD]{Spherical Harmonic Domain}
\acro{SIE}{System Identification Error}
\acro{SII}{Speech Intelligibility Index}
\acro{SIImod}{Speech Intelligibility Index in the modulation domain}
\acro{SIM}{Subscriber Identity Module}
\acro{SIMO}{Single-Input-Multiple-Output}
\acro{SINAD}{Signal-to-Noise and Distortion Ratio}
\acro{SIP}{Session Initiation Protocol}
\acro{SIR}{Signal-to-Interference Ratio}
\acro{SIR2}[SIR]{Sequential Importance Resampling}
\acro{SIREAC}{Simulation of REal Acoustics\acroextra{ Software Tool}}
\acro{SIS}{Sequential Importance Sampling}
\acro{SL}{Speech Level}
\acro{SLAM}{Simultaneous Localization and Mapping}
\acro{SLLN}{Strong Law of Large Numbers}
\acro{SLM}{Sound Level Meter}
\acro{SMA}{Spherical Microphone Array}
\acro{SMARD}{Single- and Multichannel Audio Recordings Database}
\acro{SMERSH}{Spatiotemporal Averaging Method for Enhancement of Reverberant Speech}
\acro{SMIR}{Spherical Microphone array Impulse Response}
\acro{SMPTE}{Society of Motion Picture and Television Engineers}
\acro{SMS}{Short Message Service}
\acro{SNR}{Signal-to-Noise Ratio}
\acro{SNR2}[SNR]{Speech-to-Noise Ratio}
\acro{SNT}{Subspace Noise Tracking\acroextra{ algorithm}}
\acro{SOBM}{STOI-optimal Binary Mask}
\acro{SONAR}{SOund Navigation And Ranging}
\acro{SOLA}{Synchronous Overlap Add\acroextra{. A method of scaling a signal in time and pitch independently.}}
\acro{SPC}{Specificity}
\acro{SPEECON}{Speech Databases for Consumer Devices}
\acro{SPHERE}{NIST SPeech Header REsources\acroextra{ software with embedded Shorten Compression}}
\acro{SPIN}{Speech Perception In Noise}
\acro{SPL}{Sound Pressure Level}
\acro{SPP}{Speech Presence Probability}
\acro{SPQA}{Speech Quality Assurance Package}
\acro{SQNR}{Signal-to-Quantization Noise Ratio}
\acro{SR}{Sparse Representation}
\acro{SR2}[SR]{Spectral Rotation}
\acro{SRA}{Statistical Room Acoustics}
\acro{SRI}{SRI International\acroextra{. Formerly Standford Research Institute}}
\acro{SRMR}{Speech-to-Reverberation Modulation Energy Ratio}
\acro{SRP}{Steered Response Power}
\acro{SRP-PHAT}{Steered Response Power with Phase Transform}
\acro{SRP-TDE}{Steered Response Power with Time Delay Estimation}
\acro{SRR}{Signal-to-Reverberation Ratio}
\acro{SRT}{Speech Reception Threshold\acroextra{ (also known as Speech Recognition Threshold)}}
\acro{SS}{Signal Subspace}
\acro{SSB}{Single Side-Band}
\acro{SSI}{Synthetic Sentence Identification}
\acro{SSL}{Sound Source Localization}
\acro{SSN2}[SSN]{Simultaneous Switching Noise}
\acro{SSN}{Speech-Shaped Noise}
\acro{SSNR}{Segmental \acs{SNR}}
\acro{SSOBM}{Stochastic \acs{SOBM}}
\acro{SSRR}{Segmental Signal-to-Reverberation Ratio}
\acro{SSW}{Staggered Spondaic Word}
\acro{STFT}{Short Time Fourier Transform}
\acro{STI}{Speech Transmission Index}
\acro{STIPA}{Speech Transmission Index for Public Address Systems}
\acro{STITEL}{Speech Transmission Index for Telecommunication Systems}
\acro{STMI}{Spectro-Temporal Modulation Index}
\acro{STNR}{\acs{NIST}'s Speech-to-Noise Ratio\acroextra{ Estimation Algorithm}}
\acro{STOI}{Short-Time Objective Intelligibility\acroextra{ measure}}
\acro{STQ}{Speech Processing, Transmission and Quality Aspects}
\acro{STSA}{Short Time Spectral Analysis}
\acro{STSA1}[STSA]{Short Time Spectral Amplitude}
\acro{SUS}{Semantically Unpredictable Sentences}
\acro{SVD}{Singular Value Decomposition}
\acro{SVM}{Support Vector Machine}
\acro{SWSOBM}{Stochastic \acs{WSOBM}}
\acro{T20}[$T_\textrm{20}$]{Reverberation Time\acroextra{ to decay by $20$ dB}}
\acro{T30}[$T_\textrm{30}$]{Reverberation Time\acroextra{ to decay by $30$ dB}}
\acro{T60}[$T_\textrm{60}$]{Reverberation Time\acroextra{ to decay by $60$ dB}}
\acro{TBM}{Target Binary Mask}
\acro{TDE}{Time Delay Estimation}
\acro{TDHS}{Time Domain Harmonic Scaling\acroextra{. A method of scaling a signal in time and pitch independently.}}
\acrodefplural{TDOA}[TDOAs]{Time-Differences-of-Arrival}
\acro{TDoA}{Time-Difference-of-Arrival}
\acrodefplural{TDoA}[TDoAs]{Time-Differences-of-Arrival}
\acro{TDT}{Tone Decay Test}
\acro{TF}{Time-Frequency}
\acro{TFS}{Temporal Fine Structure}
\acro{TFGM}{Time-Frequency Gain Modification\acroextra{. An approach to signal enhancement in which a signal is multiplied by a gain function in the time-frequency domain.}}
\acro{THD}{Total Harmonic Distortion}
\acro{TI}{Texas Instruments, Inc.}
\acro{TIMIT}{\acs{TI}-\acs{MIT} speech corpus}
\acro{TIPHON}{Telecommunication and Internet Protocol Harmonization Over Networks}
\acro{TLS}{Total Least-Squares}
\acro{TN}{True Negative}
\acro{TNR}{True Negative Rate}
\acro{TOA}{Time-of-Arrival}
\acrodefplural{TOA}[TOAs]{Times-of-Arrival}
\acro{TOF}{Time-of-Flight}
\acro{TOSQA}{Telekom Objective Speech Quality Assessmentt}
\acro{TP}{True Positive}
\acro{TP2}[TP]{Trivial Pursuit}
\acro{TPCC}{Trivial Pursuit with Clipping Constraints}
\acro{TPR}{True Positive Rate}
\acro{TSE}{Taylor Series Expansion}
\acro{TVAR}{Time-varying Autoregression}
\acro{UAV}{Unmanned Aerial Vehicle}
\acro{UDP}{User Datagram Protocol}
\acro{UFRJ}{Federal University of Rio de Janeiro}
\acro{UHF}{Ultra High Frequency}
\acro{UKF}{Unscented Kalman Filter}
\acro{ULA}{Uniform Linear Array}
\acro{ULF}{Ultra Low Frequency}
\acro{UMTS}{Universal Mobile Telecommunications Service}
\acro{US}{United States}
\acro{UTBM}{Universal Target Binary Mask}
\acro{VAD}{Voice Activity Detector}
\acro{VBR}{Variable Bit-Rate}
\acro{VCV}{Vowel-Consonant-Vowel}
\acro{VRD}{Variance of Decay-rates}
\acro{VGC}{Voice Grade Channel}
\acro{vMF}{von Mises-Fisher}
\acro{VoIP}{Voice Over Internet Protocol}
\acro{VRT}{Vlaamse Radio- en Televisieomroeporganisatie\acroextra{. (Flemish Radio and Television Broadcasting Organization)}}
\acro{VSELP}{Vector Sum-excited Linear Prediction}
\acro{VST}{Virtual Studio Technology\acroextra{. An interface standard developed by Steinberg for adding plugins to an audio editor.}}
\acro{WADA}{Waveform Amplitude Distribution Analysis}
\acro{WASPAA}{Workshop on Applications of Signal Processing to Audio and Acoustics}
\acro{WAV}{Waveform Audio File Format}
\acro{WAVE}{Waveform Audio File Format}
\acro{WB}{Wideband}
\acro{WER}{Word Error Rate}
\acro{WGN}{White Gaussian Noise}
\acro{WLAN}{Wireless \acs{LAN}}
\acro{WLLN}{Weak Law of Large Numbers}
\acro{WM}{Working Memory}
\acro{WMA}{Windows Media Audio}
\acro{WNG}{White Noise Gain}
\acro{WSS}{Weighted Spectral Slope}
\acro{WSOBM}{Weighted \acs{SOBM}}
\acro{WSTOI}{Weighted \acs{STOI}}
\acro{ZOS}{Zero-Order Statistics}

%% file: body.tex
\input{_intro}
\input{_baselines}
\input{_proposed}
\input{_evaluation}
\input{_conclusion}

%% file: _intro.tex
\section{Introduction}
\label{sec:intro}

With growing popularity of microphone arrays in devices such as wearables, speech enhancement takes advantage of multi-channel processing by exploiting the signal spatial characteristics. Multi-channel speech enhancement has applications in hearing aids, augmented reality, teleconferencing and robot audition \cite{Doclo2010,Lollmann2017,HaebUmbach2019} and typically consists of a \ac{MISO} block optionally followed by a single-channel post-processing. The focus in the work is on the \ac{MISO} block for a single target, however, the problem can be extended to multi-target by repeating a method for different targets.

The existing \ac{MISO} methods can be generally grouped into analytical or \ac{ML} approaches. The \ac{ML} methods \cite{Liu2018,Erdogan2016} have shown promising results but may not generalise to unseen conditions. For wearable arrays, this is potentially problematic due to the additional dimensions of complexity and variation caused by rapid movements (translation and rotation) of the array.

For analytical approaches, beamforming has been widely used for decades due to its computational simplicity and robustness. Signal-independent beamformers, such as superdirective \cite{Bitzer2001a} or delay-and-sum, provide fast and robust computation with pre-calculated weights whereas adaptive beamformers, such as \ac{MVDR} \cite{Trees2002,Capon1969}, can potentially provide better results but at the cost of more computation and the risk of signal distortion \cite{Cox1973a,Ehrenberg2010} due to errors in the array steering vector or target \ac{DOA}. A particular challenge for adaptive beamforming in the context of wearable arrays is that the short-term stationarity assumption \cite{Gannot2017} may be violated during head rotations, especially for anisotropic noise fields.

In this work, we consider the `cocktail party' \cite{Cherry1953} scenario where the subject wearing the array and the single target are surrounded by temporally dynamic ambient noise such as babble noise and with possible presence of nearby interference(s). The target \ac{DOA} with respect to the rotated array and the array's \acp{ATF} with realistic accuracy are assumed to be either known a \textit{priori} or else can be estimated \cite{Zhang2019a,Gannot2017,Schwartz2016b}.

The remainder of the paper is structured as follows: Section~\ref{sec:baselines} reviews the technical background for the signal-independent and adaptive beamformers used as baseline. Section~\ref{sec:proposed} introduces the proposed method and its novel blocks. In Section~\ref{sec:eval}, two versions of the proposed method are compared with the baseline using real-recording `cocktail party' scenario with head-worn array. Finally, conclusions are given in Section~\ref{sec:conclusion}.

%% file: _baselines.tex
\section{Baseline Beamformers}
\label{sec:baselines}

Let $\beamin\TF=[\beaminnum_{1}\TF\ldots\beaminnum_{\nMic}\TF]^{T}\in\mathbb{C}^{\nMic\times1}$ denote the vector of the observed signals $\beaminnum_{\mic}\TF$ at time frame index $\t$, frequency index $\f$, microphone index $\mic$ for a total of $
\nMic$ microphones. The beamformer output is
\begin{equation}
    \beamoutnum\TF = \weight^{H}\TF\beamin\TF,\label{eq:beamforming}
\end{equation}
where $\weight\in\mathbb{C}^{\nMic\times1}$ is beamformer weights and $(.)^H$ is the Hermitian transpose. For notational simplicity, the $\TF$ dependency will be omitted for the remaining of the paper unless stated otherwise.

Based on the \ac{MVDR} beamformer, the weights can be derived as \cite{Capon1969}
\begin{equation}
    \weight = (\ATF^{H}\NCM^{-1}\ATF)^{-1}\NCM^{-1}\ATF,\label{eq:mvdr}
\end{equation}
where $\ATF=\steervec(\doa_{s})\in\mathbb{C}^{\nMic\times1}$ is the steering vector $\steervec$ for the target \ac{DOA} $\doa_{s}$, $\NCM\in\mathbb{C}^{\nMic\times\nMic}$ is the \ac{NCM} and $(.)^{-1}$ denotes the inversion operator.

\subsection{Iso-MVDR (Superdirective)}
\label{ssec:Iso}

The Isotropic-\ac{MVDR}, referred to as Iso, assumes a stationary spherically isotropic \ac{NCM} as $\NCM$ in \eqref{eq:mvdr}. This is equivalent to assuming uncorrelated plane waves with equal power arriving from all directions. The spherically isotropic diffuse covariance matrix can be obtained as
\begin{equation}
	\NCM_{\gamma} = \int_{\doa} \steervec(\doa) {\steervec}^{H}(\doa) d\doa, 
\label{eq:theoretical_iso}
\end{equation}
where $\int_{\doa} d\doa= \int_{0}^{2\pi} \int_{0}^{\pi} \sin(\inc) d\inc d\az$ denotes integration along azimuth $\az\in[0,2\pi)$ and inclination $\inc\in[0,\pi]$.

Assuming the \ac{ATF} of the array is available for a discrete set of directions $\mathcal{I}$ from a grid of uniform distribution across azimuth and inclination, then \eqref{eq:theoretical_iso} is approximated by quadrature-weighting the grid of discrete points as
\begin{equation}
   \NCM_{\text{Iso}} =
\sum_{i \in \mathcal{I}}w_{i}\steervec(\doa_{i}) {\steervec}^{H}(\doa_{i}),
\label{eq:iso}
\end{equation}
where $w_{i}$ is the quadrature weight for each sample point given by \cite{Driscoll1994}
\begin{equation}
   w_{i} = \frac{2\sin \inc_i}{N_\az N_\inc}
\sum_{m=0}^{0.5N_\inc-1} \frac{\sin\left(\left( 2m+1\right) \inc_i \right)}{2m+1},
\label{eq:quadrature_weights}
\end{equation}
in which $\inc_i$ is the inclination of sample point $i$ and  the number of sample points in azimuth and inclination are $N_\az$ and $N_\inc$ respectively. Note that the quadrature weighting is done to preserve the uniform power isotropy by compensating for higher density of points closer to the poles in a uniform grid spatial sampling scheme. The use of other spatial sampling schemes may require no or different weighting. Using $\NCM=\NCM_{\text{Iso}}$ in \eqref{eq:mvdr} and substituting it in \eqref{eq:beamforming}, the output of this beamformer is denoted as $\beamoutnum_{\text{Iso}}$.

\subsection{MPDR}
\label{ssec:MPDR}

The \ac{MPDR} assumes \ac{SCM} as the $\NCM$ in \eqref{eq:mvdr} given as 
\begin{equation}
    \NCM_{\text{SCM}} =\expectOp\{\beamin\beamin^{H}\},\label{eq:SCM}
\end{equation}
where $\expectOp\{.\}$ is the expectation operator. An estimate of the \ac{SCM} is obtained by applying an \ac{EMA} to the instantaneous covariance matrix 
\begin{align}
    \NCM_{\text{MPDR}}\TF=\alpha\NCM_{\text{MPDR}}(\t-1,\f)\label{eq:mpdr} \\ +    (1-\alpha)\beamin\TF\beamin^{H}\TF,\nonumber
\end{align}
where $\alpha=e^{-\Delta t / T}$ is the smoothing factor (between $0$ and $1$), $\Delta t$ is the time step between frames and $T$ is the time constant.

%% file: _proposed.tex
\section{Proposed Method}
\label{sec:proposed}

\begin{figure}[t]
\centering
\includegraphics[width=.5\textwidth]{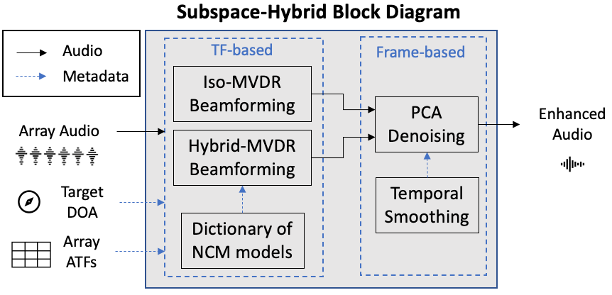}
\caption{The proposed system block diagram.}
\label{fig:system}
\end{figure}

As illustrated in Fig.~\ref{fig:system}, the proposed method consists of two-stage multi-channel processing. In the first stage two types of beamforming are performed at every \ac{TF} bin. In the second stage, the spectrum output of both beamformers at each time frame are combined to form a two-channel data on which \acs{PCA} \cite{Jolliffe2002} is performed to output the final enhanced signal. In addition to Iso-\ac{MVDR}, described in Section \ref{ssec:Iso}, the system makes use of two novel processing blocks described as follows.

\subsection{Hybrid-MVDR}
\label{ssec:Hybrid}

This block, referred to as Hybrid, performs multiple \ac{MVDR} beamforming with the same steering vector for a known target direction but various \acp{NCM} taken from a dictionary of pre-defined noise sound field models. The dictionary denoted as $P=\{\weight_{m}(\f,\Omega)\}$ contains the pre-calculated beamformer weights for each sample noise field model $m$, frequency index $\f$ and steering direction $\Omega$ where $1\leq m\leq\nModel$ for a total number of  $\nModel$ models. The output with the minimum power among the models is selected and denoted as 
\begin{align}
    & \beamoutnum_{\text{Hyb}}= (\weight_{j})^{H}\beamin,\\
    \mathrm{with\quad} & j = \argmin_{m}{\{\lVert (\weight_{m})^{H}\beamin \rVert^{2}\}}.\label{eq:minimisation}
\end{align}

The dictionary can contain beamformer weights based on a variety of noise field models such as isotropic, anisotropic, \acp{PW}, spatially uncorrelated (diagonal covariance), and potentially more complex ones such as combination of basic ones or previously measured \acp{NCM}. The size and the models used in the dictionary are described in Section~\ref{ssec:pool}.

As will be shown in Section~\ref{sec:eval}, although Hybrid beamformer results in stronger acoustic noise reduction, the output contains `musical noise' due to rapid switching of beam pattern caused by potential selection of highly different models for neighbouring time frames and frequencies. To suppress this musical noise, which is assumed to be uncorrelated, or only partially correlated, with the acoustic noise, the next proposed block extracts that component of Hybrid-\ac{MVDR} output which is correlated with the Iso-\ac{MVDR} output.

\subsection{Spectral PCA Denoising}
\label{ssec:PCA}

In each time frame, let $\beamout(\t)=[\beamoutnum(\t,1)\ldots\beamoutnum(\t,\nFreq)]^{T}$ denote the spectrum output for a beamformer with a total of $\nFreq$ frequency bands. The associated outputs from Hybrid and Iso beamformers (identified by subscript) are joint to form a two-channel complex-value array of data
\begin{equation}
    \outarray(\t) = [\beamout_{\text{Hyb}}(\t), \beamout_{\text{Iso}}(\t)].\label{eq:array_constuct}
\end{equation}
The $2\times2$ inter-channel covariance matrix of $\outarray(\t)$ is then
\begin{equation}
    \NCM(\t) =\expectOp\{\outarray^{H}(\t)\outarray(\t)\},\label{eq:Rz_theoretical}
\end{equation}
which can be approximated, using \ac{EMA}, as
\begin{equation}
    \NCM_{\text{Z}}(\t) =\alpha\NCM_{\text{Z}}(\t-1)+(1-\alpha)(\outarray^{H}(\t)\outarray(\t)).\label{eq:Rz}
\end{equation}
Using \ac{EVD}, $\NCM_{\text{Z}}$ is decomposed as
\begin{equation}
    \NCM_{\text{Z}}(\t) = \eigvec(\t)\eigval(\t)\eigvec^{-1}(\t),
\end{equation}
where $\eigvec\in\mathbb{C}^{2\times2}$ and $\eigval$ are respectively the eigenvectors and diagonal matrix of eigenvalues. Assuming the columns of $\eigval$ are sorted in descending order of eigenvalues, the first column of $\mathbf{U}(\t)=[\mathbf{U}_{S}(\t),\mathbf{U}_{N}(\t)]$ is considered as signal eigenvector denoted as $\mathbf{U}_{S}(\t)\in\mathbb{C}^{2\times1}$. 

The \ac{SS} of the $\outarray(\t)$ is reconstructed as 
\begin{equation}
    \outarray_{\text{SS}}(\t) = \outarray(\t)\eigvec_{S}(\t)\eigvec_{S}^{H}(\t),\label{eq:SS} 
\end{equation}
where the first column of $\outarray_{\text{SS}}(\t)=[\beamout_{\text{SS-Hyb}}(\t),\beamout_{\text{SS-Iso}}(\t)]$ is considered as the final spectrum output of the system and denoted as $\beamout_{\text{SS-Hyb}}$.

\subsection{Dictionary composition}
\label{ssec:pool}

\begin{figure}[t]
    \centering
    \includegraphics[scale=.7]{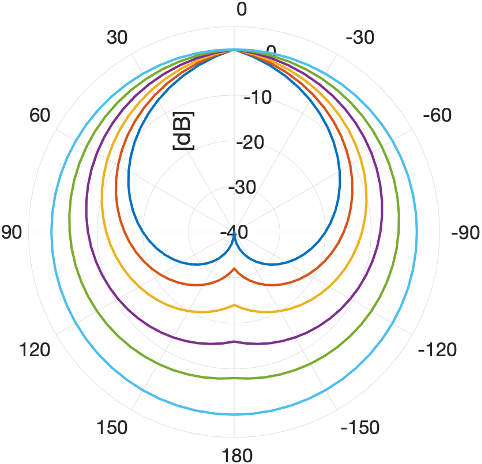}
    \caption{Isotropic and five unimodal anisotropic models for the horizontal isotropy of the noise field.}
    \label{fig:isotropy}
\end{figure}

Two variations of dictionary $P$, in Hybrid are considered using available \acp{ATF}. In the first version, denoted as SS-Hyb, the \ac{NCM} models consist of the identity matrix, spherically isotropic noise and five unimodal anisotropic distributions across horizon, as shown in Fig.~\ref{fig:isotropy}, horizontally rotated for every $\SI{6}{\deg}$ azimuth spacing as for the peak position and were quadrature weighted along the inclination to form the 3D sound field. For unimodal anisotropic models, the power was a linear function of azimuth with power dynamic ranges of $\{8,16,24,32,40\} \SI{}{\dB}$. The second version, denoted as SS-HybX, extends the number of models in the dictionary by additionally including individual \ac{PW} models for all available \ac{ATF} directions.

%% file: _evaluation.tex
\section{Evaluations}
\label{sec:eval}

\begin{figure*}[t]
    \centering
    \includegraphics[width=\textwidth]{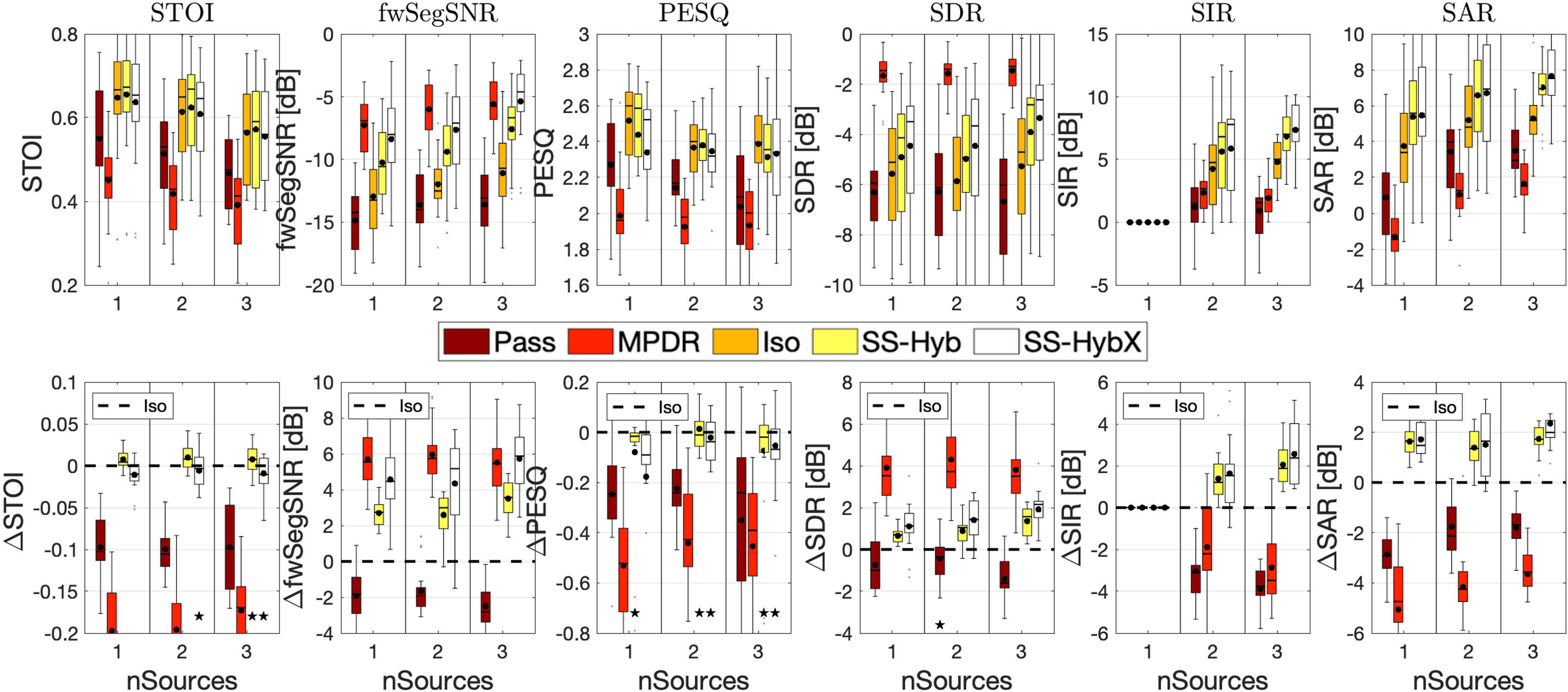}
    \caption{The absolute (top) and relative (bottom) STOI \cite{Taal2011}, fwSegSNR \cite{Hu2008b}, PESQ \cite{Rix2001}, SDR \cite{vincentPerformanceMeasurementBlind2006a}, SIR \cite{vincentPerformanceMeasurementBlind2006a} and SAR \cite{vincentPerformanceMeasurementBlind2006a}.}
    \label{fig:results}
\end{figure*}

In this section, the two implementations of the proposed method are compared with the Iso-\ac{MVDR} and adaptive \ac{MPDR} baseline beamformers as well as the `passthrough' signal at the reference microphone. 

\subsection{Dataset and Array}
\label{ssec:dataset}
In the context of augmented hearing and augmented reality, the EasyCom dataset \cite{Donley2021a} was used for evaluation. It contains recordings of `cocktail-party' scenarios where a subject with 6-channel head-worn array (four microphones fixed to a pair of glasses and two positioned in the ears) was sat down with multiple talkers at a table surrounded by ten loudspeakers playing restaurant-like ambient noises. The target \ac{DOA} over time is provided via head-tracking metadata for each talker. For this evaluation, the dataset is split into $\SI{6}{\s}$ chunks such that the target onset occurs after $\SI{2}{\s}$ based on the voice activity metadata in \cite{Donley2021a}. For the results visualization, the chunks were categorized according to the number of active sources per chunk denoted as `nSources' = $[1, 2, 3]$. Close-talking headset microphones for each talker, also included in \cite{Donley2021a}, were time- and level-aligned to the array reference microphone for use in intrusive metrics.

To avoid spatial aliasing due to the array geometry, data was down-sampled to $\SI{10}{\kHz}$ sample rate. The \ac{STFT} used $\SI{16}{\ms}$ time-window and $\SI{8}{\ms}$ step. The smoothing factor $\alpha$ in \eqref{eq:mpdr} and \eqref{eq:Rz} was chosen empirically with $T=\SI{50}{\ms}$ and $T=\SI{80}{\ms}$, respectively, for
\ac{MPDR} and SS-Hybrid. The condition number of the $\NCM$ for PWs in the dictionary of SS-HybX was limited to maximum of $100$ via \acs{NCM} diagonal loading to avoid ill-condition covariance matrices. Our investigation showed no necessity of condition number limiting for the other \ac{NCM} models, at least for this dataset.

\subsection{Results and Discussions}
\label{ssec:dataset}

\begin{figure}[t]
    \centering
    \includegraphics[scale=.32]{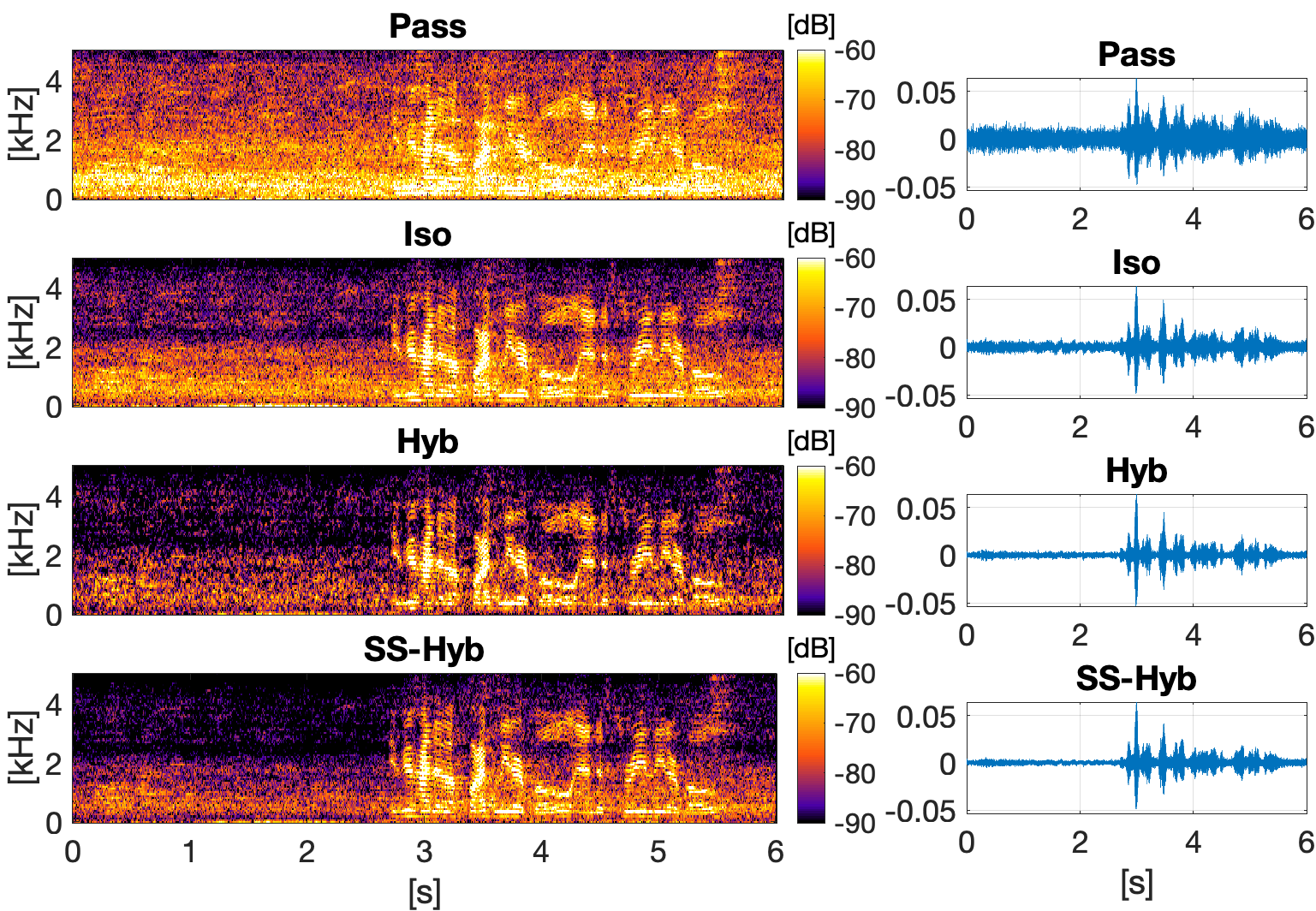}
    \caption{Spectrograms of the methods for a trial (nSources=$1$).}
    \label{fig:stft}
\end{figure}

Some audio examples of the results and animated visualization of the beam patterns are available at \cite{Demo}. Figure~\ref{fig:results} shows the absolute (top row) and relative (bottom row) performance according to various intrusive metrics. The black dot indicates the mean while a star indicates no significant difference from the Iso-MVDR (dashed line) according to paired t-test at $p=5\%$ level. 

It can be seen that SS-Hyb outperforms the best baseline (Iso) by an average of $0.01$ STOI, $\SI{3}{\dB}$ fwSegSNR, $\SI{1.5}{\dB}$ SDR, $\SI{2}{\dB}$ SIR, $\SI{1.8}{\dB}$ SAR while sharing the best PESQ with Iso. On the other hand, SS-HybX performs better than SS-Hyb in noise suppression with additional mean of $\SI{2}{\dB}$ fwSegSNR, $\SI{1}{\dB}$ SDR, SIR and SAR particularly in the presence of interferer talker(s) (nSource$>1$), due to utilization of \ac{PW} models, while sharing the same STOI and PESQ with Iso. Although \ac{MPDR} provides the highest noise suppression, substantial target distortion caused by proneness to imperfection in target \acp{ATF} and \ac{DOA} leads to poor STOI, PESQ, SIR and SAR scores.

Figure~\ref{fig:stft} shows the spectrograms of Passthrough, Iso, Hybrid and SS-Hybrid for a representative trial. Although Hybrid provides more noise suppression than Iso, it contains noticeable musical noise, which is suppressed in SS-Hybrid via \ac{PCA}. 

%% file: _conclusion.tex
\section{Conclusions}
\label{sec:conclusion}

A novel two-stage multi-channel speech enhancement method is proposed which combines the robustness and computational simplicity of signal-independent beamforming with the performance of adaptive beamforming. The system proposes a Hybrid beamformer which performs multiple \ac{MVDR} beamformers with different noise field models including isotropic and a variety of anisotropic distributions. The outputs from an Iso and optimal Hybrid beamformers are then jointly used to remove the musical noise via multi-channel \ac{PCA} denoising. The evaluation results, using real-recording `cocktail-party' scenario with head-worn microphone array, demonstrate the benefit of the proposed method in term of improved noise suppression and speech intelligibility with similar speech quality metric, compared to the baseline static and adaptive beamformers.